\documentclass[11pt,a4paper]{article}

\renewcommand{\author}{Emil Khalisi}
\newcommand{\paperlabel}{E.\ Khalisi \& J.\ Gripp (2020)}
\newcommand{\titel}{Hierarchical Eclipses}

\renewcommand{\date}{\today}

\usepackage{hyperref}
\hypersetup{pdfauthor={Emil Khalisi}}
\hypersetup{pdftitle={\titel}}

\usepackage[T1]{fontenc}        
\usepackage{mathptmx}           
\usepackage{pifont}             
\usepackage[scaled=0.92]{helvet} 
\usepackage[british]{babel}     
\usepackage{amsmath}            
\usepackage{microtype}          
\usepackage{multicol}           
\usepackage{multirow}           
\usepackage{eurosym}            
\usepackage{units}              
\usepackage{setspace}           

\usepackage{calc}
\usepackage[a4paper]{geometry}  
\usepackage{scrextend}           
\changefontsizes{10pt}

\usepackage{titlesec}
\titleformat*{\section}{\large\bfseries}
\titleformat*{\subsection}{\normalsize\bfseries}

\usepackage{graphicx}           
\usepackage{float}              
\usepackage{caption}            
\usepackage{fancyhdr}
\renewcommand{\headrulewidth}{0.4pt}
\usepackage{colortbl}             
\definecolor{grey20}{RGB}{208,208,208}

\begin{document}


\fancyhead{}
\fancyhead[LO]{%
   \footnotesize \textsc{German Version published in:} \\
   {\footnotesize \textit{Sternzeit 45},
      No.\ 2 / 2020, p82--91 (ISSN: 0721-8168)}
}
\fancyhead[RO]{
   \footnotesize {\tt arXiv:2005.07131 [astro-ph.EP]}\\
   \footnotesize {v2: 30th May 2020}%
}
\fancyfoot[C]{\thepage}

\renewcommand{\abstractname}{}

\twocolumn[                
\begin{@twocolumnfalse}    

\section*{\centerline{\LARGE \titel }}

\begin{center}
{\author , Joachim Gripp \\}
\textit{Sternzeit e.V., Heidelberg and Kiel, Germany}\\
\textit{e-mail:} \texttt{khalisi }$\;$ or $\;$ \texttt{gripp$\;$ \dots @sternzeit-online.de}
\end{center}

\vspace{-\baselineskip}
\begin{abstract}
\changefontsizes{10pt}

\noindent
\textbf{Abstract.}
The obscuration of a celestial body that covers another one in the
background will be called a ``hierarchical eclipse''.
The most obvious case is that a star or a planet will be hidden
from sight by the moon during a lunar eclipse.
Four objects of the solar system will line up then.
We investigate this phenomenon with respect to the region of
visibility and periodicity.
There exists a parallax field constraining the chances for
observation.
A historic account from the Middle Ages is preserved that we
analyse from different viewing angles.
Furthermore, we provide a list of events from 0 to 4000 AD.
From this, it is apparent that Jupiter is most often involved
in such spectacles because its orbit inclination is small.
High-inclination orbits reduce the probability to have a
coincidence of an occultation of that object with a lunar eclipse.

\vspace{\baselineskip}
\noindent
\textbf{Keywords:}
Occultations,
Lunar Eclipses,
Celestial Mechanics,
Planets,
Solar System

\vspace{\baselineskip}
\noindent
{\small Received: 6 February 2020. Accepted: 4 March 2020}

\end{abstract}

\centerline{\rule{0.8\textwidth}{0.4pt}}
\vspace{1.5\baselineskip}

\end{@twocolumnfalse}   
]                       



\enlargethispage{0.5ex}

\section{Introduction}

Occultations of planets in the solar system by the moon receive
particular attention in astronomical almanacs.
They occur at semi-regular intervals of a few years' time and,
as such, they are quite common.
In the 21st century there are a total of 59 cases for the classical
naked-eye planets, 34 of which happen during daytime.
Lunar eclipses are more frequent, and for a fixed location one can
expect roughly 1 per year on average.

The impressive lunar eclipse of 27 July 2018 took place in the
vicinity of Mars.
This gave rise to the idea of a combination of both spectacles:
a ``hierarchical eclipse''.
It will be characterised by four bodies placed in a straight line
--- Sun, Earth, Moon, and a planet.
In this paper we put the ``double overlaps'', as they would be seen
from the sun, to a test.
We try to gain insight into the following questions:
Did this kind of consecutive covering of celestial objects happen
in the historical past?
In particular, are there reports about an eclipsed moon eclipsing
another planet simultaneously?
How often does such a line-up occur?
When will the next opportunity be scheduled?
Do exist cycles or, at least, accumulations for these hierarchical
eclipses?

The analysis is based on an empirical list of events compiled for
the years from 0 to +4,000 AD.
We used the simulation software packages \textit{Guide 9.1} (2017),
\textit{Cartes du Ciel 4.0} (2017), and \textit{Occult 4.6.0}
(2018).
The next section presents some few accounts from history that match
our configuration.
In Section \ref{ch:parallax} we check the tolerance for visibility,
and in the subsequent section we discuss the frequency of the
occurrences.
Within the scope of our investigation, we achieved some
instructive results.

For the sake of clarity, we use the following technical terms:
The obscuration of a planet will be called ``occultation'' in
order to distinguish it from the ``eclipse'' of the moon.
The instant of disappearance of the planet behind the moon's disk is
labelled ``immersion'', while its re-appearance is the ``emersion''.
The eclipse contacts with the earth's umbra are assigned U1 to U4,
as defined in the astronomical textbooks:
begin of the umbral eclipse (U1), begin of totality (U2), end of
totality (U3), and end of the declining phase (U4), respectively.

\section{Records from History}

One historic account about a hierarchical eclipse was passed down
by the medieval priest Roger of Hoveden.
His lifetime can only be limited to the years between 1174 to
1201 AD.
In the relevant paragraph of his chronicle he reports about a
``bright star'' being involved in a peculiar scene with the
eclipsed moon.
The record for the year 756 AD reads as follows \cite{rogerdehov}:
\begin{quote}
On the eighth day before the calends of December (23rd November),
the moon, on her fifteenth day, being about her full, appeared
to be covered with the colour of blood,
and then, the darkness decreasing, she turned to her usual
brightness; but, in a wondrous manner, a bright star followed the
moon, and passing across her, preceded her when shining,
at the same distance at which it had followed her before she was
darkened.
%
\end{quote}

\begin{figure}[t]
\centering
\includegraphics[width=\linewidth]{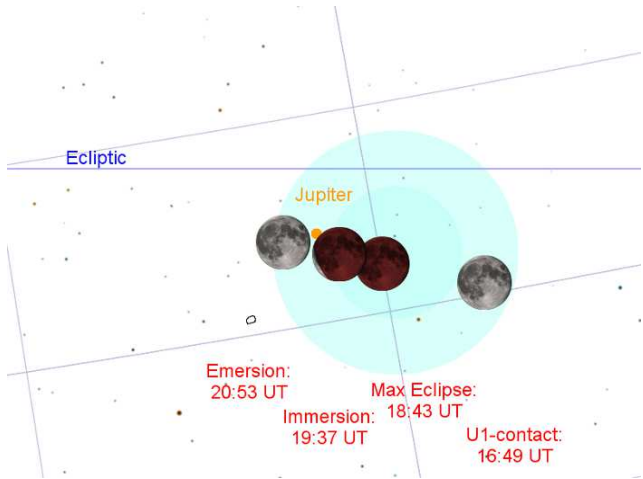}
\caption{Jupiter was occulted by the lunar disk, which suffered an
   eclipse itself, on 23 November 755 AD.
   Simulated view from London.}
\label{fig:jupitermofi}
\end{figure}

The procedure is not correctly described from the astronomical
point of view.
The direction of the course is inverted, and the lunar eclipse
took place one year earlier, in 755 AD.
The ``star'' was Jupiter who was overtaken by the moon while the
latter was darkened in the earth's shadow
(Figure \ref{fig:jupitermofi}).
The event goes back more than four centuries prior to Roger's
time.
It appears only natural that scribal errors slip in, especially,
when one of the copists in the chain of transmission failed to
understand the meaning of the line himself.
The original record does not seem to be preserved.

%
\renewcommand{\headrulewidth}{0pt}
\fancyhead{}
\fancyhead[CE, CO]{\footnotesize \itshape \paperlabel : \titel}

Another record, that would be contemporary with Roger's life,
closely missed a similar sight in Syria.
The patriarch of the Orthodox Church in Aleppo, Michael Syrus
(1126--1199), vividly depicts the nightmare of a solar eclipse
on 11 April 1176.
Thereafter he continues \cite{histeclipses}:
\begin{quote}
Fifteen days after [the solar eclipse], in this month of Nissan
(April) at the decline of Monday, at dusk, there was an eclipse
of the Moon in the part of the sky where the eclipse of the Sun
had taken place \dots\
%
\end{quote}

Of course, it took place in the opposite part of the sky.
The eclipse was partial (mag = 0.673) with the Northern rim
unobscured.
Above the non-eclipsed edge, Jupiter was shining.
The cleric must have regarded the bright spot as a usual star and
did not mention it.
Somewhat further to the South that spot was occulted, e.g.\ in
Kenya and, even more, in South Africa.

Another great occasion to watch a hierarchical eclipse occurred
shortly before the telescopic era.
On 26 July 1580 the eclipsed moon met even two planets, Saturn and
Uranus.
An observer in Japan could have seen the disappearance of Saturn,
while someone in Northern Australia the same of Uranus.
Only on the small Indonesian island of Koror both planets were
occulted, though not simultaneously.
The partially eclipsed moon ran over Uranus first, and 10 minutes
after its emersion, the occultation of Saturn began.
The subsequent lunation provided an occultation of both planets at
the same time, but without the eclipse then.
However, in that year no-one knew about the existence of Uranus,
and telescopes were not invented yet.

As regards modern times, the sole chance to have had a glimpse of
Saturn being covered during an eclipse was on 14 December 1796.
It was visible in East Asia, but any remark about an observation
is not known to us.

\section{The View from Earth} \label{ch:parallax}

The occultation of any background object during a lunar eclipse
implies its opposition with the sun.
As a matter of fact, Mercury and Venus are excluded.

At first, let us imagine a fictitious observer in the center of
the earth looking through a vitreous sphere.
The shadow of the earth (umbra) has an apparent radius as large as
1.3$^{\circ}$ at the distance of the moon.
The geocentric observer would see the maximum duration when the
planet is placed exactly in the ecliptic and the moon traverses
the shadow centrally.
Then we have the longest totality, and the moon crosses over the
planet alongside its diameter, also providing the longest
occultation possible.
Such a configuration almost never happens in practice.
Usually, the moon passes through the shadow at some displacement
from the center, and the planet will hide along a cord behind the
moon's face.

\begin{figure}[t]
\centering
\includegraphics[width=\linewidth]{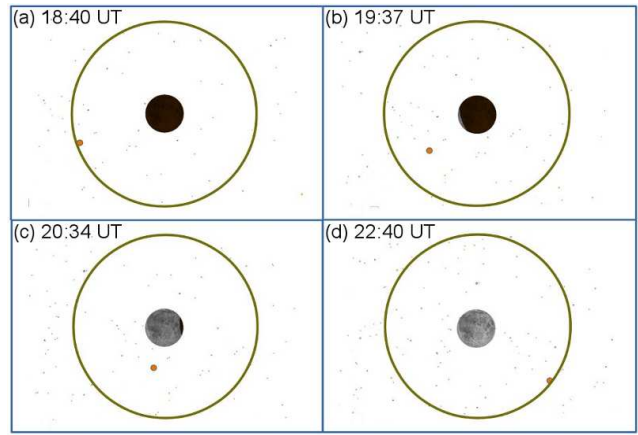}
\caption{Geocentric views of Jupiter and Moon during the lunar
    eclipse of 755 at four different times.
    The circle denotes the parallactic field of view for the
    occultation to be visible from an unspecified geographic
    location somewhere on Earth.
    Compare Figure \ref{fig:geokarte}.}
\label{fig:parafield}
\end{figure}

For the other extreme, the moon just scratches both the planet
and the umbra with its diametrically opposite fringes.
On one edge it would cover the shining dot for a moment and,
on the other, it touches the umbra on its antipode.
This adds an angle of 0.45$^{\circ}$ to 0.52$^{\circ}$ to the
extent of the terrestrial shadow,
depending on the current distance of the moon, since the size of
its disk varies between perigee and apogee.
For the planet, it gives $\approx 1.8^{\circ}$ of tolerance to
stay above or below the ecliptic (Figure \ref{fig:parafield}).

\begin{figure*}[t]
\centering
\includegraphics[width=\linewidth]{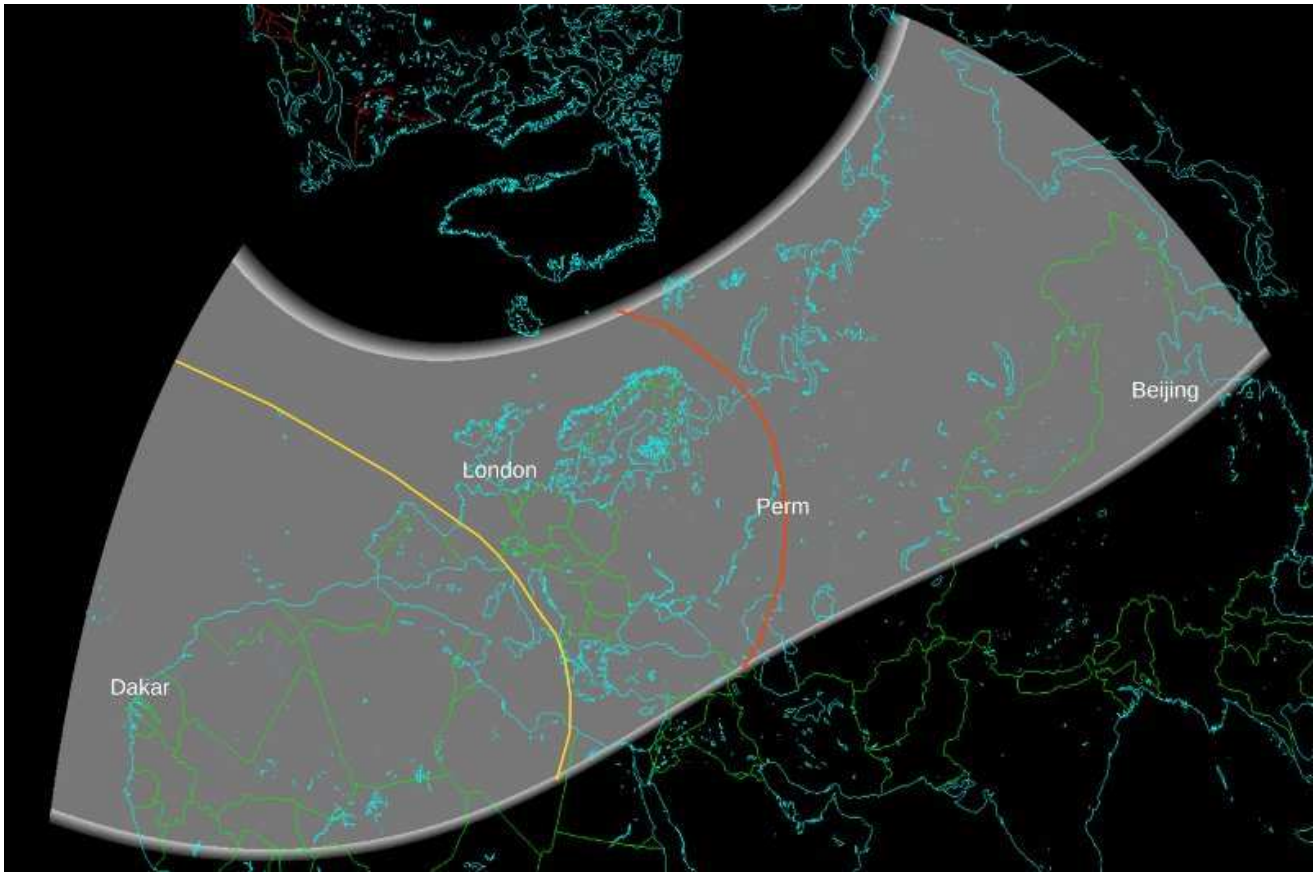}
\caption{Region of visibility for the Jupiter occultation of
    755 AD.}
\label{fig:geokarte}
\vspace{-0.5\baselineskip}
\end{figure*}

In contrast to that geocentric observer, the real viewer has the
advantage to move on the surface of the earth.
His topocentric position gains a parallax, as it would be seen
from the moon (cosine of his geographical latitude).
At the poles this accounts for another $\approx 15^{\prime}$.
On the whole, we find the parallactic field of view to be 4 times
larger than the diameter of the moon.
To ensure the hierarchical eclipse to be seen from some spot on
earth (topocentric), the planet must stay inside this parallactic
circle (geocentric).
If the planet is outside, there will be no point on the globe for
the hierarchical eclipse.

Figure \ref{fig:parafield} shows the field of tolerance for
Roger's occultation event.
From the geocentric view, four instants are presented:
\begin{itemize}
\item[(a)] At 18:40 UT, the circle reached Jupiter.
On the surface of the earth, both the planet and the moon were
rising in Dakar, West Africa.
The planet was standing already in the umbra, and the occultation
of the planet (its immersion) could be seen during totality there.
However, from the location of London, Jupiter resided in the
penumbra, while the moon was still running towards the planet.
\item[(b)] At 19:37 UT, the geocentric observer would have seen
the planet closer to the darkened face, while in London its
immersion occurred.
But the totality has already ended at 19:25 UT, such that the
occultation started during the decreasing partial phase.
Jupiter entered the umbra, while covered, and returned for
visibility (emersion) after the lunar eclipse had fully ended,
i.e.\ beyond the U4-contact.
\item[(c)] Going even further eastward, the occultation of Jupiter
would be observed from another angle, e.g.\ in Perm in the Ural:
immersion as well as emersion took place with Jupiter standing in
the penumbra.
The terrestrial shadow crept up slowly towards the planet, while
the moon was overtaking it.
When the moon reached the planet, the eclipse was close to finish
(penumbra neglected).
\item[(d)] For Beijing it was not until 22:40 UT as the moon caught
up with the shining dot, and the whole procedure passed off
sequentially instead of simultaneously.
\end{itemize}

Note that the fictitious observer at the center of the earth
would not have seen any occultation at all.
It is a pure effect of the extent of the spherical earth that
the hierarchical eclipse happened.
The region of visibility is shown as the grey-shaded area in
Figure \ref{fig:geokarte}.
East of the yellow line the total eclipse ended (U3-contact),
and the red line marks the border of the finish of the partial
phase (U4-contact).

A useful insight is that there are no fixed times for immersion
and emersion, but they depend on the location of the observer.
The occultation occurs at different times in spite of having
deployed the same time frame for reference (like the UT).
With regard to Earth's shadow, the planet seems to stay at
different positions in the sky.
While the eclipse is in progress, the umbra of the earth moves
on, too, reducing the deviation for an eastward observer.

\vspace{-\baselineskip}
\noindent
\begin{table*}[t!]
\vspace*{-0.5\baselineskip}
\caption{Eclipsed moon occults a planet as visible for a ``good'' observing site.
    Times in [UT].}
\label{tab:luneclipseplanet}
\centering
\begin{tabular}{|rl|cc|ccc|l|}
\hline
\rowcolor{grey20}
\multicolumn{2}{|c|}{\cellcolor{grey20} Date [AD]}
                & Ecl. U1 -- U4  & Magn.& Planet  & Bri. &  I  --  E      & Best visibility \\
\hline
     2 & Nov  8 & 21:26 -- 23:59 & 0.46 & Mars    & -1.8 & 21:58 -- 23:02 & W-Brazil \\
   195 & Jul 10 & 23:41 -- 03:36 & 1.71 & Saturn  &  0.6 & 03:35 -- 04:42 & Central Pacific \\
   354 & Dec 16 & 12:28 -- 15:50 & 1.33 & Saturn  & -0.4 & 15:27 -- 16:22 & E-Africa \\
   400 & Dec 17 & 17:05 -- 20:27 & 1.06 & Jupiter & -2.7 & 19:46 -- 21:11 & Seychelles \\
   412 & Nov  4 & 18:22 -- 22:06 & 1.60 & Mars    & -1.9 & 20:06 -- 21:05 & S-Africa \\
   458 & Nov  6 & 21:16 -- 00:11 & 0.80 & Jupiter & -2.8 & 21:10 -- 22:16 & Caucasus \\
   480 & Sep  5 & 03:25 -- 06:19 & 0.60 & Jupiter & -2.9 & 03:21 -- 03:58 & Hudson-Bay \\
*  502 & Dec 29 & 12:52 -- 16:34 & 1.64 & Saturn  & -0.3 & 13:24 -- 14:29 & Hawaii \\
   513 & Jun  4 & 08:02 -- 11:50 & 1.34 & Jupiter & -2.7 & 07:23 -- 08:39 & Bolivia \\
   524 & May  3 & 16:25 -- 20:02 & 1.65 & Jupiter & -2.6 & 19:59 -- 20:37 & S-Pole \\
   755 & Nov 23 & 16:49 -- 20:37 & 1.40 & Jupiter & -2.8 & 19:37 -- 20:53 & Europe \\
   771 & Feb  4 & 08:29 -- 11:32 & 0.93 & Saturn  &  0.3 & 11:27 -- 12:25 & Tasmania \\
   799 & Jul 21 & 13:47 -- 17:30 & 1.56 & Jupiter & -2.9 & 17:02 -- 18:17 & Kazakhstan \\
*  810 & Jun 20 & 18:04 -- 21:36 & 1.84 & Jupiter & -2.8 & 20:11 -- 21:34 & Madagascar \\
   821 & May 20 & 18:30 -- 22:14 & 1.41 & Jupiter & -2.6 & 21:42 -- 22:52 & E-Brazil \\
   879 & Apr 10 & 09:29 -- 12:52 & 1.36 & Jupiter & -2.5 & 12:08 -- 12:43 & S-Pole \\
   959 & Jun 23 & 06:37 -- 09:44 & 0.94 & Saturn  &  0.1 & 09:42 -- 10:30 & Antarctica \\
  1052 & Dec  8 & 20:38 -- 00:07 & 1.65 & Jupiter & -2.7 & 23:56 -- 00:53 & Caribbean \\
  1176 & Apr 25 & 17:43 -- 20:46 & 0.67 & Jupiter & -2.5 & 19:20 -- 20:29 & Indian Ocean \\
  1234 & Mrc 17 & 02:06 -- 04:51 & 0.65 & Jupiter & -2.5 & 02:55 -- 03:47 & Patagonia \\
  1312 & Jun 19 & 18:05 -- 21:06 & 1.55 & Saturn  &  0.1 & 20:05 -- 21:11 & Namibia \\
  1407 & Nov 15 & 11:05 -- 14:25 & 1.19 & Jupiter & -2.8 & 11:16 -- 11:44 & N-Siberia \\
  1418 & Oct 14 & 20:15 -- 23:53 & 1.12 & Jupiter & -2.9 & 22:40 -- 22:48 & N-Canada \\
  1462 & Jun 12 & 00:32 -- 03:17 & 0.59 & Jupiter & -2.7 & 00:16 -- 00:58 & Antarctica \\
* 1473 & May 12 & 06:21 -- 08:27 & 0.37 & Jupiter & -2.6 & 07:09 -- 08:26 & Polynesia \\
  1531 & Apr  1 & 18:05 -- 19:24 & 0.11 & Jupiter & -2.5 & 19:01 -- 20:03 & S-Africa \\
  1580 & Jul 26 & 09:27 -- 12:47 & 1.26 & Saturn  &  0.3 & 10:52 -- 11:58 & Japan (+ Uranus!) \\
  1591 & Dec 30 & 02:12 -- 05:45 & 1.57 & Saturn  & -0.4 & 05:14 -- 06:11 & Alaska \\
  1796 & Dec 14 & 13:05 -- 15:27 & 0.49 & Saturn  & -0.3 & 14:52 -- 15:55 & Siberia \\
  2344 & Jul 26 & 10:40 -- 14:21 & 1.34 & Saturn  &  0.1 & 12:18 -- 13:44 & N-Pacific \\
  2429 & Jun 17 & 10:42 -- 11:10 & 0.02 & Saturn  &  0.1 & 10:58 -- 11:58 & New Zealand \\
  2488 & Apr 26 & 07:42 -- 11:02 & 1.38 & Mars    & -1.6 & 08:07 -- 08:51 & Antarctica \\
  2829 & Jan 11 & 01:41 -- 05:33 & 1.81 & Saturn  & -0.4 & 05:26 -- 06:40 & N-Pacific \\
  2932 & Jun  9 & 22:13 -- 23:50 & 0.21 & Jupiter & -2.6 & 22:32 -- 23:38 & E-Brazil \\
* 2977 & Jan 26 & 07:03 -- 10:49 & 1.65 & Saturn  & -0.4 & 07:37 -- 08:48 & S-Mexico \\
  2990 & May  1 & 23:57 -- 01:09 & 0.09 & Jupiter & -2.5 & 00:26 -- 01:21 & S-Chile \\
  3108 & Jun 15 & 06:26 -- 08:57 & 0.44 & Saturn  &  0.0 & 06:17 -- 07:47 & French Polynesia \\
  3218 & Jul 30 & 00:00 -- 02:23 & 0.46 & Jupiter & -2.7 & 00:06 -- 01:05 & Madagascar \\
  3229 & Jun 28 & 16:30 -- 19:20 & 0.55 & Jupiter & -2.6 & 18:03 -- 18:48 & Madagascar \\
  3287 & May 19 & 14:30 -- 17:32 & 0.88 & Jupiter & -2.5 & 14:36 -- 15:27 & Fr.\ S-Antarctica \\
  3376 & Aug 21 & 18:58 -- 20:39 & 0.20 & Saturn  &  0.2 & 19:06 -- 20:33 & Maldives \\
  3444 & Dec 17 & 16:17 -- 19:07 & 0.59 & Saturn  & -0.4 & 17:13 -- 17:30 & Arctic \\
  3461 & Jul 14 & 22:35 -- 02:12 & 1.62 & Saturn  &  0.0 & 22:10 -- 22:41 & Antarctica \\
  3584 & Jun  6 & 12:52 -- 16:32 & 1.16 & Jupiter & -2.5 & 14:26 -- 14:58 & Madagascar \\
  3805 & Jan 28 & 08:12 -- 11:19 & 0.93 & Jupiter & -2.7 & 08:29 -- 09:17 & Central Chile \\
  3815 & Dec 29 & 12:28 -- 15:37 & 0.89 & Jupiter & -2.9 & 13:19 -- 13:58 & Svalbard \\
  3826 & Nov 27 & 15:22 -- 18:25 & 0.65 & Jupiter & -2.9 & 17:40 -- 18:36 & Svalbard \\
  3870 & Jul 26 & 00:20 -- 03:57 & 1.48 & Jupiter & -2.7 & 00:33 -- 01:34 & Seychelles \\
  3881 & Jun 25 & 08:28 -- 12:04 & 1.82 & Jupiter & -2.6 & 09:26 -- 10:10 & S-Australia \\
\hline
\end{tabular}
\end{table*}

\section{Frequency of Hierarchical Eclipses}

\begin{figure*}[t]
\vspace*{-0.5\baselineskip}
\centering
\includegraphics[width=\linewidth]{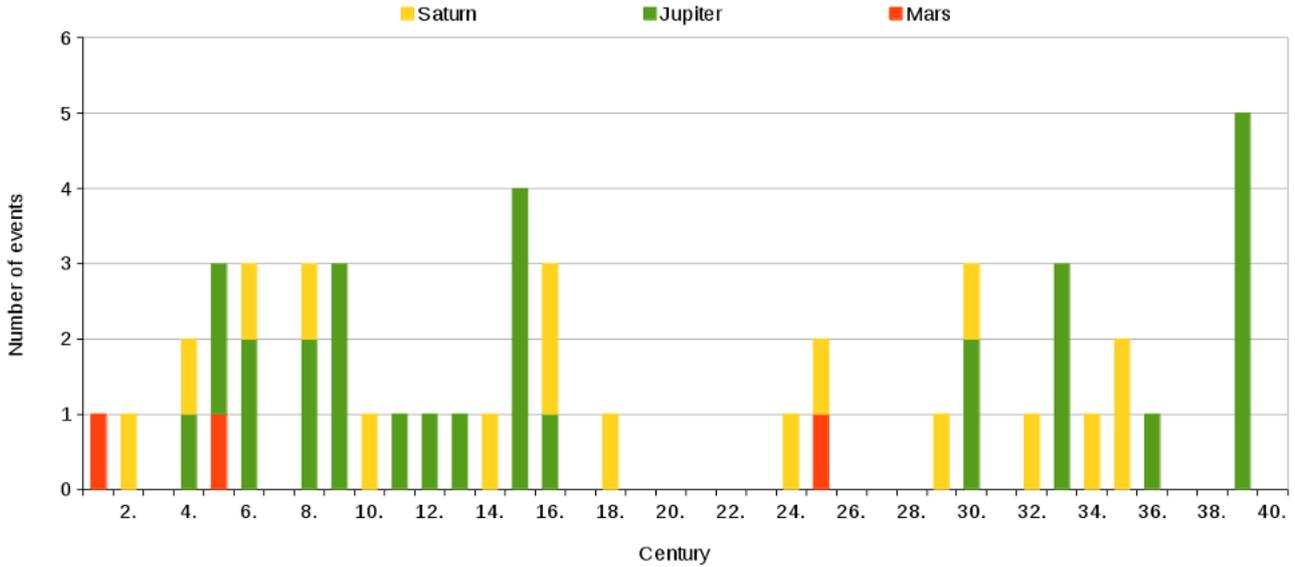}
\caption{Number of hierarchical eclipses per century.}
\label{fig:haeufigkeit}
\vspace{-0.5\baselineskip}
\end{figure*}

Table \ref{tab:luneclipseplanet} lists all hierarchical eclipses
we were able to find between 0 and 4,000 AD.
We make no claim for completeness.
Dates before 1582 are Julian, thereafter Gregorian.
Considered are only cases with at least one planetary contact
(immersion or emersion) inside the time interval for the eclipse,
which is given in column 2: between U1 and U4.
The magnitude in column 3 denotes the maximum eclipse.
If below 1.0, the eclipse is partial.
Columns 4, 5, and 6 give the name of the planet, its brightness,
and the time of immersion (I) and emersion (E).
These latter times correspond to a ``good'' site of visibility in
the last column 7.
It may not necessarily be an ideal spot, but it would be very
close to it.
The star (*) in the first column indicates the hierarchical
coverage for the fictitious position at the center of the earth.

It is informative to discover that only four (*) of 49 incidents
could be seen from the geocentric point.
The others are attributed to the extended surface of the earth.
That is to say, we observe them, \emph{because} one is placed at
a certain parallax somewhere on the globe.
As the case would be, one could witness either the eclipsing
hierarchy, or a simple occultation of the planet, or even nothing.
The lunar eclipse, though, is visible in the same way for anyone
having the moon above the horizon.

Figure \ref{fig:haeufigkeit} shows a histogram of the data for
each century.
Hierarchical eclipses seem to be irregularly distributed.
There are intervals of accumulation, but we live in an
extraordinary long interval of shortage.
Jupiter was prominent in the first millennium, while Mars is very
rarely occulted, in general.
A strict periodicity cannot be extracted for any planet, but some
repetitions and gaps do catch attention.
The reasons for it rest upon the characteristics of the planetary
orbits, as Meeus \textit{etal.} point out \cite{meeus-etal_1977}:
inclination and eccentricity.
The period of eclipses, which is governed by the draconitic
period of the moon, is also essential.

For obtaining a cycle, three periods need to be considered.
The synodic month of the moon has to be an integer, otherwise
there is no full moon.
The draconitic month has to be half its number, otherwise there
will be no eclipse.
And, thirdly, the synodic revolution of the planet must be an
integer, too, in order to meet the opposition.
All periods are incommensurable and carry a minute displacement
against the other on the long run.
Here we sketch the qualities briefly for each planet, but a more
precise mathematical treatment is still pending.

\paragraph{Mars}
has an orbital inclination of 1.85$^{\circ}$ to the ecliptic,
but it exhibits the largest departure at the extremes, as seen
from Earth.
There are only two small windows, each of about 50$^{\circ}$
centered on the nodes with the ecliptic, in which the planet
crosses this reference plane within a favourable latitude
accessible for a lunar eclipse.
So, the hierarchical eclipse can only happen in the days of
April/May and from October to the beginning of December.
For the other months the ecliptic latitude of Mars will be too
high or too low, and, thus, out of reach for the parallactic
field.

A secondary obstacle regards the ellipticity of its orbit.
The velocity will not be uniform owing to the perihelion and
aphelion.
Therefore, this gives rise to an advance or retard as compared to
a circular orbit.
This means that the lunar node could fail to catch the planet at
a convenient instant although the conditions are fulfilled.
The ``period'' for a recurrence, if it exists, will possibly be
valid for a small piece of its orbit only, unless extremely long
time scales are envisioned.

\paragraph{Jupiter}
displays the smallest deviation from the ecliptic and can be
occulted by an eclipsed moon in any season.
This planet is most often involved in hierarchical eclipses,
as Figure \ref{fig:haeufigkeit} confirms.
However, there are episodes of abundances as well as paucities.
Two slight periods of 10.9 and 57.9 years flash up, and they
\emph{would} be more prominent, if the orbit was circular.
On the other side, these two quasi-periods have to be merged with
the advancing difference with respect to the lunar node.
The revolution of the lunar node is controlled by the precession
of the moon's plane, and should comply with half its number of
the draconitic period which is 18.61 years.

Again, both series above only satisfy the conditions for a small
arc of Jupiter's orbit.
If the incident comes about close to perihelion, then the 11-year
period can hold for one or two more chances as in 799--810--821.
When the minute differences on subsequent ``hits'' have
accumulated, the series tears off and there are no hierarchical
eclipses for several centuries.
See \cite{meeus-etal_1977} for details.

\paragraph{Saturn}
is an intermediate case.
The inclination of its orbit is 2.49$^{\circ}$ but the planet's
outward distance makes the vertical elevation from the ecliptic
appear $\pm2.3^{\circ}$ at maximum.
For a few weeks in spring and autumn the planet is beyond the
threshold for eclipses, and the hierarchy is suspended.
An asset is that Saturn changes its position along the orbit quite
slowly and stays inside the admissible belt for lunar eclipses for
several years.
The conditions seem to provide a larger stability, but we confess
not having checked this in detail.
We were not able to identify a period for Saturn, for all its
hierarchical eclipses seem to proceed at random.

\paragraph{The Moon}
itself is liable to the extent of the circular parallactic field.
Its elliptical orbit around Earth brings the anomalistic month as
another period into consideration.
This type rules the exact size of the circle and determines whether
or not the planet will be positioned inside or just slightly outside
the ring.
The circle ``pulsates'' in the rhythm of the anomalistic month.

\vspace{\baselineskip}
For extremely long timescales, the eccentricity of each orbit
varies as well.
This holds even for the earth itself.
Taken all these factors into account, there will hardly be a
cycle of a stable nature.
Anyway, all periods turn out incommensurable in the Solar System,
while the purpose of any search for periods is usually to find an
approximation as good as possible.

\section{Anti-Transits}

In closing, let us change perspectives.
If the moon is able to cover a background object, then the sun
will do so as well.
Taking planets as targets, they will hide behind the solar disk.
One may call this an ``anti-transit''.
As a matter of principle, it is unobservable, and our examination
becomes just an academic question.

In order to have this state of affairs hierarchically, the sun
needs to be covered, too, i.e.\ the moon will be the obvious
object to trigger a solar eclipse while the planet is
anti-transiting.
To achieve that, the planet has to be in the superior conjunction
and stay close to one of its nodes.
The maximum ecliptic latitude allowed is $\pm15^{\prime}$, since
this is the apparent radius of the sun.
Mercury and Venus can join our consideration again.
The duration of the anti-transit depends on the relative speed
between the sun and the planet.
The sun traverses its own diameter in about 12 hours, however,
Mercury and Venus move faster than the sun at their superior
conjunction and may stay longer than a day behind it, if the
passage is central.

If you think, it is too weird, you're wrong.
It happens from time to time, most recently at the total
eclipse of 30 June 1954 \cite{ricci}.
Jupiter stepped behind the sun at 10 UT that day and remained
obscured for the next 17 hours.
The moon entered the playground around midday (Southern
Scandinavia) and caused an eclipse of the sun between 11 and 14 UT.
Thus, two celestial objects were deprived from sight for the
observer on Earth.
The next opportunity is envisaged for 14 May 2105, when Mercury
will perform its anti-transit and a partial solar eclipse will
have its stage.

\section{Conclusions}

We presented the rare phenomenon of an ``eclipse-oc\-cul\-ta\-tion''
when a planet at opposition is eclipsed by the moon which,
in turn, is eclipsed by the earth.
The example of Roger de Hoveden showed that Jupiter's disappearance
occurs at different times for different places, though the same time
frame is used.
This effect is due to the parallax for the observers on the surface
of the earth.
In contrast to that, the lunar eclipse for all observers occurs at
the same instant.

As known since Antiquity, lunar eclipses can be utilised to measure
the time difference between two places on Earth, if they are widely
spaced in geographical longitude.
In fact, this method was employed in old times for localising time
zones or synchronising clocks.

Occultations by the moon can be used to determine its position
and speed in the sky.
Hence, they reveal the secular acceleration that is based on the
exchange of angular momentum between Earth and Moon.
For an evaluation of those observations and results, the geographical
position of the observer is relevant.
The method fails to work when documents from various (unknown)
cultural regions are compared, because the occultation takes place
at different times, even if a common time scale like the ``UT''
is used.
In case of known places of observation, still a correction
procedure has to be applied.

\section*{Acknowledgments}

We thank Dr.\ Eckhard Fliege for critically reading the manuscript.
The results of this paper were published in the popular German
magazine for astronomy, \textit{Sternzeit 45}, No.\ 2 / 2020.
They are also included in the Habilitation (ch.\ 7.7) by EK
submitted to the University of Heidelberg, Germany, in February
2020.


\end{document}